# Semiquantum private comparison via cavity QED

Xin Xu, Jiang-Yuan Lian, Tian-Yu Ye*

College of Information & Electronic Engineering, Zhejiang Gongshang University, Hangzhou 310018, P.R.China

E-mail：yetianyu@zjgsu.edu.cn (T.Y. Ye)

**Abstract:** In this paper, we design the first semiquantum private comparison (SQPC) protocol which is realized via cavity quantum electrodynamics (QED) by making use of the evolution law of atom. With the help of a semi-honest third party (TP), the proposed protocol can compare the equality of private inputs from two semiquantum parties who only have limited quantum capabilities. The proposed protocol uses product states as initial quantum resource and employs none of unitary operations, quantum entanglement swapping operation or delay lines. Security proof turns out that it can defeat both the external attack and the internal attack.

**Keywords:** Semiquantum private comparison (SQPC); Cavity quantum electrodynamics (QED); Two-atom product state

## 1 Introduction

Differing from the classical cryptography, quantum cryptography can gain the unconditional security theoretically with the aid of laws of quantum mechanics including the quantum no-cloning theorem and the Heisenberg uncertainty theorem and so on. Since Bennett and Brassard [1] proposed the first quantum key distribution (QKD) protocol in 1984, quantum cryptography has aroused the interests of scholars all over the world. With the development of quantum cryptography, it has produced plenty of branches, such as QKD [1-3], quantum secret sharing (QSS) [4-5], quantum secure direct communication (QSDC) [6-7], quantum signature (QS) [8-9], and so on. In the past two decades, a new application of quantum cryptography named as quantum private comparison (QPC) has also been developed greatly [10-13]. QPC can be traced back to the classical millionaire problem proposed by Yao [14] in 1982, where two millionaires want to figure out who is richer without knowing each other's actual property.

Semiquantumness means that not all participants have full quantum capabilities, in other words, at least one participant can only perform limited quantum operations. In a semiquantum cryptography scheme, the semiquantum participant is only allowed to deliver qubits, generate qubits in the classical basis, measure qubits with the classical basis and scramble qubits. [15-17] Up to now, great progress has been achieved on semiquantum private comparison (SQPC) so that numerous SQPC protocols based on different quantum states and different quantum technologies have been proposed, such as those in Refs.[18-34] *etc*.

Cavity quantum electrodynamics (QED) is one of the most popular physical systems for quantum information processing. The high level of coherence of cavity QED makes it ideal to realize the quantum information processing experiment. [35] In 2017, Ye [36] proposed the first QPC scheme via cavity QED. At present, there are no SQPC protocols which are realized via cavity QED. Inspired of the protocol in Ref.[36], by making use of the evolution law of atom, we put forward the first SQPC protocol which is realized via cavity QED. The proposed protocol uses product states as initial quantum resource and employs none of unitary operations, quantum entanglement swapping operation or delay lines.

## 2 Model of cavity QED

After two equal two-level atoms enter into a single-mode cavity, they interact with it motivated by a classical field. The interaction Hamiltonian within the rotating-wave approximation in this circumstance is [37-40]

$$H=\omega_0 S_z+\omega_a a^\dagger a+\sum_{j=1}^{2}[g\left(a^\dagger S_j^-+aS_j^\dagger\right)+\Omega\left(S_j^\dagger e^{-i\omega t}+S_j^- e^{i\omega t}\right)]. \tag{1}$$

Here, $\omega_0$ is the atomic transition frequency, $\omega_a$ is the cavity frequency, $a^\dagger$ is the creation operator, $a$ is the annihilation operator, $g$ is the atom-cavity coupling strength, $\Omega$ is the Rabi frequency, $\omega$ is the classical field frequency, $t$ is the interaction time, $S_z=(1/2)\sum_{j=1}^{2}\left(|e_j\rangle\langle e_j|-|g_j\rangle\langle g_j|\right)$, $S_j^-=|g_j\rangle\langle e_j|$ and $S_j^\dagger=|e_j\rangle\langle g_j|$, where $|g_j\rangle$ and $|e_j\rangle$ are the ground state and the excited state of the $j$ th atom, respectively. When $\omega=\omega_0$, within the interaction picture, the system evolution operator is [37-40]

$$U(t)=e^{-iH_0 t}e^{-iH_e t}, \tag{2}$$

where $H_e$ is the effective Hamiltonian and $H_0=\Omega\sum_{j=1}^{2}\left(S_j^\dagger+S_j^-\right)$. When $\delta\gg g$ and $\Omega\gg\delta,g$, the influences of cavity decay and thermal field are eliminated, where $\delta$ is the detuning between $\omega_0$ and $\omega_a$. Accordingly, $H_e$ becomes [37-40]

$$H_e=(\lambda/2)\left[\sum_{j=1}^{2}\left(|e_j\rangle\langle e_j|+|g_j\rangle\langle g_j|\right)+\sum_{i,j=1,i\neq j}^{2}\left(S_i^\dagger S_j^-+S_i^\dagger S_j^\dagger+H.C.\right)\right], \tag{3}$$

where $\lambda=g^2/2\delta$. When $\lambda t=\pi/4$ and $\Omega t=\pi$, two atoms undergo the evolution and become entangled:

$$|gg\rangle_{jk}\to\frac{1}{\sqrt{2}}e^{-i\pi/4}\left(|gg\rangle_{jk}-i|ee\rangle_{jk}\right), \tag{4}$$

$$|ge\rangle_{jk}\to\frac{1}{\sqrt{2}}e^{-i\pi/4}\left(|ge\rangle_{jk}-i|eg\rangle_{jk}\right), \tag{5}$$

$$|eg\rangle_{jk}\to\frac{1}{\sqrt{2}}e^{-i\pi/4}\left(|eg\rangle_{jk}-i|ge\rangle_{jk}\right), \tag{6}$$

$$|ee\rangle_{jk}\to\frac{1}{\sqrt{2}}e^{-i\pi/4}\left(|ee\rangle_{jk}-i|gg\rangle_{jk}\right). \tag{7}$$

Let

$$\left|\phi^-\right\rangle_{jk} = \frac{1}{\sqrt{2}}\left(\left|gg\right\rangle_{jk} - i\left|ee\right\rangle_{jk}\right), \tag{8}$$

$$\left|\psi^-\right\rangle_{jk} = \frac{1}{\sqrt{2}}\left(\left|ge\right\rangle_{jk} - i\left|eg\right\rangle_{jk}\right), \tag{9}$$

$$\left|\psi^+\right\rangle_{jk} = \frac{1}{\sqrt{2}}\left(\left|eg\right\rangle_{jk} - i\left|ge\right\rangle_{jk}\right), \tag{10}$$

$$\left|\phi^+\right\rangle_{jk} = \frac{1}{\sqrt{2}}\left(\left|ee\right\rangle_{jk} - i\left|gg\right\rangle_{jk}\right) . \tag{11}$$

Apparently, $\Delta = \left\{\left|\phi^-\right\rangle_{jk}, \left|\psi^-\right\rangle_{jk}, \left|\psi^+\right\rangle_{jk}, \left|\phi^+\right\rangle_{jk}\right\}$ form a complete orthogonal basis for four-dimensional Hilbert space.

## 3 The proposed SQPC Protocol

Suppose that there are two semiquantum participants, Alice and Bob. Alice and Bob possess the private sequences $M_A = \left\{M_A^1, M_A^2, \ldots, M_A^L\right\}$ and $M_B = \left\{M_B^1, M_B^2, \ldots, M_B^L\right\}$, respectively, where $M_A^j, M_B^j \in \left\{0,1\right\}$, $j = 1,2,\ldots,L$ and $L$ is the length of $M_A$ or $M_B$. Moreover, they pre-share a secret key $K_{AB} = \left\{K_{AB}^1, K_{AB}^2, \ldots, K_{AB}^L\right\}$ with a secure SQKD protocol [41], where $K_{AB}^j \in \left\{0,1\right\}$ and $j = 1,2,\ldots,L$. They intend to know the equality of their private sequences under the help of a semi-honest third party (TP) who possesses full quantum abilities. The term 'semi-honest' signifies that TP is allowed to misbehave arbitrarily but cannot conspire with anyone else. [12] The proposed SQPC protocol is described concretely as follows, whose flow chart is shown in Fig.1.

**Step 1:** TP prepares $8L$ product states, each of which is randomly chosen from one of the four states $\left\{\left|gg\right\rangle, \left|ge\right\rangle, \left|eg\right\rangle, \left|ee\right\rangle\right\}$. These product states form a quantum state sequence which is depicted as $S = \left[S^1, S^2, \ldots, S^{8L}\right]$, where $S^i \in \left\{\left|gg\right\rangle, \left|ge\right\rangle, \left|eg\right\rangle, \left|ee\right\rangle\right\}$ and $i = 1,2,\ldots,8L$.

**Step 2:** TP transmits $S^i$ into the single-mode cavity illustrated in Sect.2, where $i = 1,2,\ldots,8L$, and makes $\Omega t = \pi$ and $\lambda t = \pi/4$. Consequently, the two atoms of $S^i$ evolve according to formulas (4)-(7) and are turned into an entangled state in the end. When they leave the single-mode cavity, TP collects them in hand. Then, TP makes the first and the second atoms from all $S^i$ after evolution to make up sequences $S_A$ and $S_B$, respectively, where $S_A = \left[S_A^1, S_A^2, \ldots, S_A^{8L}\right]$ and $S_B = \left[S_B^1, S_B^2, \ldots, S_B^{8L}\right]$. Finally, TP transmits the atoms of $S_A$ to Alice and the atoms of $S_B$ to Bob one by one. Note that TP sends out the next atom only after obtaining the previous one, except the first one.

**Step 3:** Alice installs a wavelength filter and a photon number splitter in front of her own devices to erase the influence of Trojan horse attacks. [42,43] When receiving $S_A^i$ from TP, Alice randomly chooses to perform the SIFT operation or the CTRL operation on it. Specifically, the SIFT operation means to measure the received atom with the $Z$ basis (i.e., $\{|g\rangle, |e\rangle\}$), record the measurement result, generate a new one in the same state as the found state and send it to TP. The CTRL operation means to reflect the received atom with no disturbance. For clarity, the sequence of $S_A$ after Alice's operations is represented as $S_A^{'}$.

Bob installs a wavelength filter and a photon number splitter in front of his own devices to prevent Trojan horse attacks. [42,43] When getting $S_B^i$ from TP, Bob also randomly chooses the SIFT operation or the CTRL operation for it. For clarity, the sequence of $S_B$ after Bob's operations is denoted as $S_B^{'}$.

**Step 4:** Alice and Bob inform TP of which atoms they have performed the SIFT operations on. TP executes the corresponding actions according to this information.

Case ①: both Alice and Bob chose the CTRL operation. TP measures $S_A^{'i}$ and $S_B^{'i}$ with the $\Delta$ basis. TP compares her measurement result with the theoretical evolution result of $S^i$ to calculate the error rate. If the error rate is unreasonably high, the protocol will be terminated; otherwise, the protocol will be continued;

Case ②: Alice chose the SIFT operation while Bob chose the CTRL operation. TP measures $S_A^{'i}$ and $S_B^{'i}$ with the $Z$ basis and asks Alice to publish her measurement result on $S_A^i$. TP judges whether the measurement result announced by Alice, the initial state of $S^i$ and her own measurement results on $S_A^{'i}$ and $S_B^{'i}$ are correctly related or not, according to formulas (4-7). If the error rate is unreasonably high, the protocol will be terminated; otherwise, the protocol will be continued;

Case ③: Alice chose the CTRL operation while Bob chose the SIFT operation. TP measures $S_A^{'i}$ and $S_B^{'i}$ with the $Z$ basis and requires Bob to publish his measurement result on $S_B^i$. TP judges whether the measurement result announced by Bob, the initial state of $S^i$ and her own measurement results on $S_A^{'i}$ and $S_B^{'i}$ are correctly related or not, according to formulas (4-7). If the error rate is unreasonably high, the protocol will be terminated; otherwise, the protocol will be continued;

Case ④: both Alice and Bob chose the SIFT operation. TP randomly selects half of the positions belonging to this case, and requires Alice and Bob to publish their corresponding measurement results on the atoms of $S_A$ and $S_B$ on these chosen positions, respectively. TP measures the atoms of $S_A^{'}$ and $S_B^{'}$ on these chosen positions with the $Z$ basis and judges whether her measurement results, the informed measurement results from Alice and Bob and the initial states of the corresponding product states in $S$ are correctly correlated or not. If the error rate is unreasonably high, the protocol will be terminated; otherwise, the protocol will be continued.

**Step 5:** Alice utilizes her measurement results on the remaining atoms of $S_A$ in Case ④ to produce the bit sequence $K_A = \{K_A^1, K_A^2, \ldots, K_A^L\}$ according to the following rule: when her measurement result on $S_A^j$ is $|g\rangle$, set $K_A^j = 0$; and when her measurement result on $S_A^j$ is $|e\rangle$, set $K_A^j = 1$. Then, Alice calculates $R_A^j = M_A^j \oplus K_A^j \oplus K_{AB}^j$, where $\oplus$ is the modulo 2 addition operation. Here, $j = 1, 2, \ldots, L$. Finally, Alice sends $R_A$ to TP, where $R_A = \{R_A^1, R_A^2, \ldots, R_A^L\}$.

Bob uses his measurement results on the remaining atoms of $S_B$ in Case ④ to generate the bit sequence $K_B = \{K_B^1, K_B^2, \ldots, K_B^L\}$ according to the following rule: when his measurement result on $S_B^j$ is $|g\rangle$, set $K_B^j = 0$; and when his measurement result on $S_B^j$ is $|e\rangle$, set $K_B^j = 1$. Here, $j = 1, 2, \ldots, L$. Then, Bob calculates $R_B^j = M_B^j \oplus K_B^j \oplus K_{AB}^j$. Finally, Bob sends $R_B$ to TP, where $R_B = \{R_B^1, R_B^2, \ldots, R_B^L\}$.

**Step 6:** TP transforms the initial state of $S^j$ corresponding to the remaining atom in Case ④ into one classical bit according to the following rule: when the initial state of $S^j$ is $|gg\rangle$ or $|ee\rangle$, set $K_C^j = 0$; and when the initial state of $S^j$ is $|ge\rangle$ or $|eg\rangle$, set $K_C^j = 1$. Here, $j = 1, 2, \ldots, L$. Then, TP computes $R^j = R_A^j \oplus R_B^j \oplus K_C^j$ in order. Once TP finds $R^j = 1$ for certain $j$, she stops calculating and concludes $M_A \neq M_B$; otherwise, she concludes $M_A = M_B$. Finally, TP tells Alice and Bob the final comparison result of $M_A$ and $M_B$ publicly.

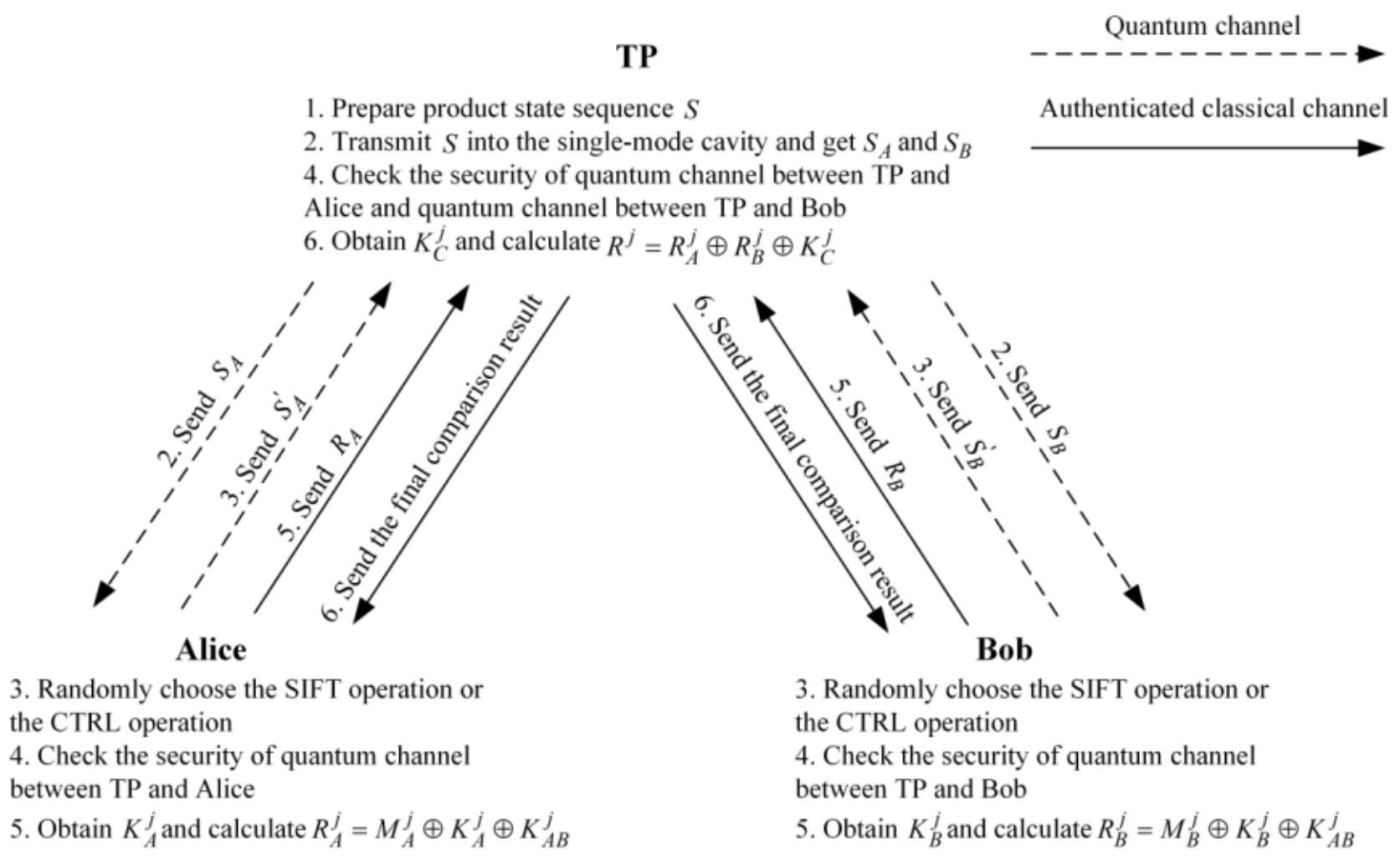

Fig.1 The flow chart of the proposed SQPC protocol

# 3 Analysis

## 3.1 Correctness Analysis

In this protocol, Alice utilizes her measurement results on the remaining atoms of $S_A$ in Case ④ to produce $K_A$, while Bob uses his measurement results on the remaining atoms of $S_B$ in Case ④ to generate $K_B$. TP transforms the initial state of $S^j$ corresponding to the remaining atom in Case ④ into $K_C^j$. It is easy to know from formulas (4-7) that $K_A^j \oplus K_B^j \oplus K_C^j = 0$. Hence, it has

$$\begin{aligned} R^j &= R_A^j \oplus R_B^j \oplus K_C^j \\ &= \left(M_A^j \oplus K_A^j \oplus K_{AB}^j\right) \oplus \left(M_B^j \oplus K_B^j \oplus K_{AB}^j\right) \oplus K_C^j \\ &= \left(M_A^j \oplus M_B^j\right) \oplus \left(K_A^j \oplus K_B^j \oplus K_C^j\right) \\ &= M_A^j \oplus M_B^j. \end{aligned} \tag{12}$$

It can be deduced out from Eq.(12) that only when $R^j = 0$ stands for $j = 1, 2, \ldots, L$ can we get $M_A = M_B$. Therefore, it is evident that the output correctness of the proposed SQPC protocol has been verified.

For clarity, we give out the corresponding relations among different parameters in the proposed SQPC protocol in Table 1 to further illustrate its output correctness, taking $K_{AB}^j = 0$ for example.

Table 1 The corresponding relations among different parameters in the proposed SQPC protocol (assume $K_{AB}^j = 0$ )

| $M_A^j$ | $M_B^j$ | $S^j$ | $K_C^j$ | $S_A^j S_B^j$ | $K_A^j$ | $K_B^j$ | $R_A^j$ | $R_B^j$ | $R^j$ |
|---|---|---|---|---|---|---|---|---|---|
| 0 | 0 | $\|gg\rangle$ | 0 | $\|gg\rangle$ | 0 | 0 | 0 | 0 | 0 |
| | | | | $\|ee\rangle$ | 1 | 1 | 1 | 1 | 0 |
| | | $\|ge\rangle$ | 1 | $\|ge\rangle$ | 0 | 1 | 0 | 1 | 0 |
| | | | | $\|eg\rangle$ | 1 | 0 | 1 | 0 | 0 |
| | | $\|eg\rangle$ | 1 | $\|ge\rangle$ | 0 | 1 | 0 | 1 | 0 |
| | | | | $\|eg\rangle$ | 1 | 0 | 1 | 0 | 0 |
| | | $\|ee\rangle$ | 0 | $\|gg\rangle$ | 0 | 0 | 0 | 0 | 0 |
| | | | | $\|ee\rangle$ | 1 | 1 | 1 | 1 | 0 |
| 0 | 1 | $\|gg\rangle$ | 0 | $\|gg\rangle$ | 0 | 0 | 0 | 1 | 1 |
| | | | | $\|ee\rangle$ | 1 | 1 | 1 | 0 | 1 |
| | | $\|ge\rangle$ | 1 | $\|ge\rangle$ | 0 | 1 | 0 | 0 | 1 |
| | | | | $\|eg\rangle$ | 1 | 0 | 1 | 1 | 1 |
| | | $\|eg\rangle$ | 1 | $\|ge\rangle$ | 0 | 1 | 0 | 0 | 1 |
| | | | | $\|eg\rangle$ | 1 | 0 | 1 | 1 | 1 |
| | | $\|ee\rangle$ | 0 | $\|gg\rangle$ | 0 | 0 | 0 | 1 | 1 |
| | | | | $\|ee\rangle$ | 1 | 1 | 1 | 0 | 1 |
| 1 | 0 | $\|gg\rangle$ | 0 | $\|gg\rangle$ | 0 | 0 | 1 | 0 | 1 |
| | | | | $\|ee\rangle$ | 1 | 1 | 0 | 1 | 1 |
| | | $\|ge\rangle$ | 1 | $\|ge\rangle$ | 0 | 1 | 1 | 1 | 1 |
| | | | | $\|eg\rangle$ | 1 | 0 | 0 | 0 | 1 |
| | | $\|eg\rangle$ | 1 | $\|ge\rangle$ | 0 | 1 | 1 | 1 | 1 |
| | | | | $\|eg\rangle$ | 1 | 0 | 0 | 0 | 1 |
| | | $\|ee\rangle$ | 0 | $\|gg\rangle$ | 0 | 0 | 1 | 0 | 1 |
| | | | | $\|ee\rangle$ | 1 | 1 | 0 | 1 | 1 |
| 1 | 1 | $\|gg\rangle$ | 0 | $\|gg\rangle$ | 0 | 0 | 1 | 1 | 0 |
| | | | | $\|ee\rangle$ | 1 | 1 | 0 | 0 | 0 |
| | | $\|ge\rangle$ | 1 | $\|ge\rangle$ | 0 | 1 | 1 | 0 | 0 |
| | | | | $\|eg\rangle$ | 1 | 0 | 0 | 1 | 0 |
| | | $\|eg\rangle$ | 1 | $\|ge\rangle$ | 0 | 1 | 1 | 0 | 0 |
| | | | | $\|eg\rangle$ | 1 | 0 | 0 | 1 | 0 |
| | | $\|ee\rangle$ | 0 | $\|gg\rangle$ | 0 | 0 | 1 | 1 | 0 |
| | | | | $\|ee\rangle$ | 1 | 1 | 0 | 0 | 0 |

### 3.2 Security Analysis

In this section, both the external attack and the internal attack are analyzed for security verification.

#### 3.2.1 External attack

As $M_A$ and $M_B$ are encrypted by $K_A$ and $K_B$, respectively, in order to steal $M_A$ and $M_B$, Eve, who is an eavesdropper, needs to get $K_A$ and $K_B$ first. Possible attacks executed by Eve include the intercept-resend attack, the measure-resend attack, the entangle-measure attack, the Trojan horse attack, *et al.* As the transmission of $S_A$ is independent from that of $S_B$, in the following, without loss of generality, we only focus on analyzing Eve's attacks on the quantum channel between TP and Alice.

**(1) Intercept-resend attack**

Eve intercepts $S_A^i$ delivered by TP and sends a fake atom generated in the $Z$ basis in advance to Alice. When both Alice and Bob choose the CTRL operation, Eve's mischievous behavior on $S_A^i$ is detected with the probability of $\frac{3}{4}$; when at least one of Alice and Bob chooses the SIFT operation, Eve's mischievous behavior on $S_A^i$ is detected with the probability of $\frac{1}{2}$. As there are $8L$ atoms in $S_A$, Eve's intercept-resend attack on $S_A$ is detected with the probability of $1-\left[1-\left(\frac{1}{4}\times\frac{3}{4}+\frac{1}{4}\times\frac{1}{2}+\frac{1}{4}\times\frac{1}{2}+\frac{1}{4}\times\frac{1}{2}\times\frac{1}{2}\right)\right]^{8L}=1-\left(\frac{1}{2}\right)^{8L}$. This probability approaches 1 with an enough large $L$.

**(2) Measure-resend attack**

Eve intercepts $S_A^i$ delivered by TP, measures it with the $Z$ basis and sends the resulted state to Alice. When both Alice and Bob choose the CTRL operation, Eve's mischievous behavior on $S_A^i$ is detected with the probability of $\frac{1}{2}$; when at least one of Alice and Bob chooses the SIFT operation, Eve's mischievous behavior on $S_A^i$ cannot be detected. As there are $8L$ atoms in $S_A$, Eve's measure-resend attack on $S_A$ is detected with the probability of $1-\left[1-\left(\frac{1}{4}\times\frac{1}{2}\right)\right]^{8L}=1-\left(\frac{7}{8}\right)^{8L}$. This probability is close to 1 with an enough large $L$.

**(3) Entangle-measure attack**

Eve's entangle-measure attack can be depicted as Fig.2, where $U_E$ and $U_F$ are two unitary operations sharing a common probe space with the initial state $|\xi\rangle_E$. [15,16] Concretely speaking, Eve launches her entangle-measure attack by performing $U_E$ on the atoms from TP to Alice and Bob and then applying $U_F$ on the atoms back from Alice and Bob to TP.

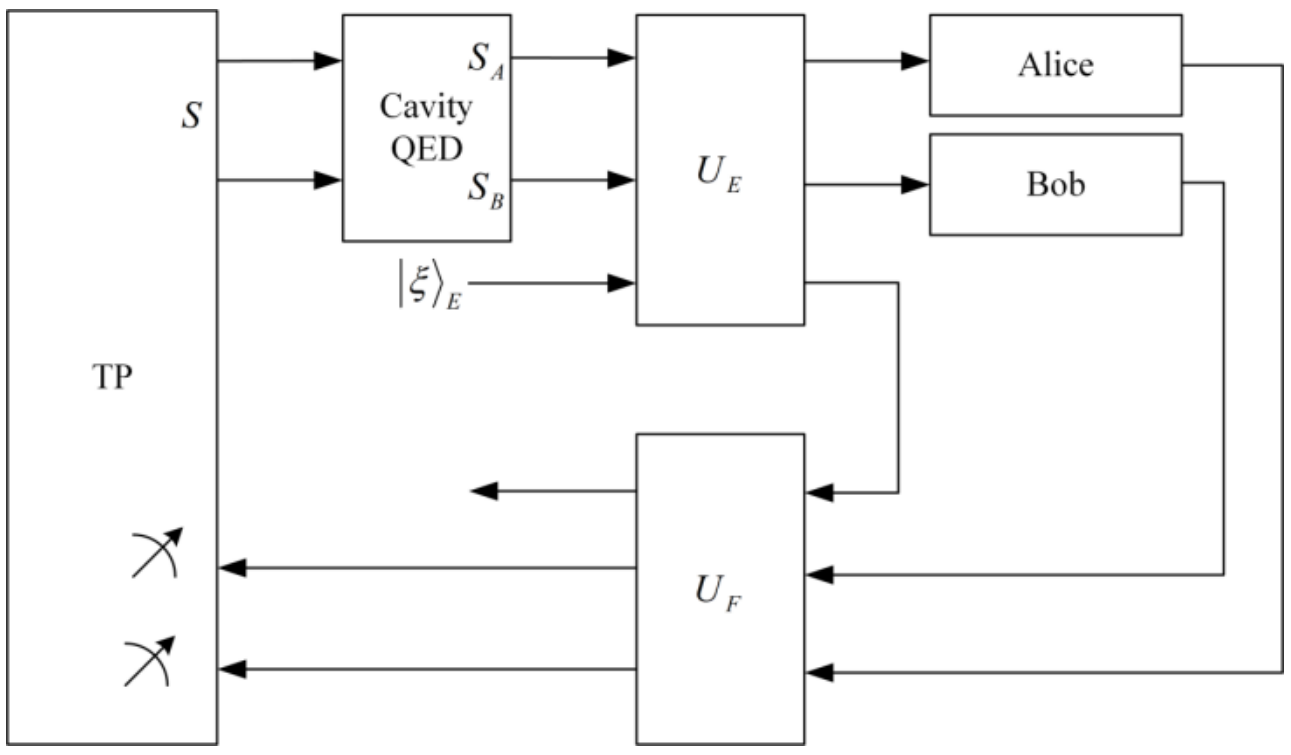

Fig.2 Eve's entangle-measure attack with $U_E$ and $U_F$

**Theorem 1.** *Suppose that Eve performs* $U_E$ *on the atoms from TP to Alice and Bob and* $U_F$ *on the atoms back from Alice and Bob to TP. In order to introduce no error in Step 4, the final state of Eve's probe should be independent of not only the operations of Alice and Bob but also their measurement results. Consequently, Eve can't get any information about* $K_A$ *or* $K_B$.

**Proof.** Without loss of generality, suppose that the initial state of $S^i$ is $|gg\rangle$. The effect of $U_E$ can be described as follows:

$$U_E\left(|g\rangle|\xi\rangle_E\right)=\beta_{gg}|g\rangle|\xi_{gg}\rangle+\beta_{ge}|e\rangle|\xi_{ge}\rangle, \tag{13}$$

$$U_E\left(|e\rangle|\xi\rangle_E\right)=\beta_{eg}|g\rangle|\xi_{eg}\rangle+\beta_{ee}|e\rangle|\xi_{ee}\rangle, \tag{14}$$

where $|\xi_{gg}\rangle$, $|\xi_{ge}\rangle$, $|\xi_{eg}\rangle$ and $|\xi_{ee}\rangle$ are probe states of Eve, $|\beta_{gg}|^2+|\beta_{ge}|^2=1$ and $|\beta_{eg}|^2+|\beta_{ee}|^2=1$.

The global state of the composite system after Eve imposes $U_E$ is

$$\begin{aligned}
&U_E\left[\frac{1}{\sqrt{2}}e^{-i\pi/4}\left(|gg\rangle_{AB}-i|ee\rangle_{AB}\right)|\xi\rangle_E\right]\\
&=\frac{1}{\sqrt{2}}e^{-i\pi/4}\Big[\left(\beta_{gg}|g\rangle_A|\xi_{gg}\rangle+\beta_{ge}|e\rangle_A|\xi_{ge}\rangle\right)\left(\beta_{gg}|g\rangle_B|\xi_{gg}\rangle+\beta_{ge}|e\rangle_B|\xi_{ge}\rangle\right)\\
&\qquad-i\left(\beta_{eg}|g\rangle_A|\xi_{eg}\rangle+\beta_{ee}|e\rangle_A|\xi_{ee}\rangle\right)\left(\beta_{eg}|g\rangle_B|\xi_{eg}\rangle+\beta_{ee}|e\rangle_B|\xi_{ee}\rangle\right)\Big]\\
&=\frac{1}{\sqrt{2}}e^{-i\pi/4}\Big[|g\rangle_A|g\rangle_B\left(\beta_{gg}^2|\xi_{gg}\rangle|\xi_{gg}\rangle-i\beta_{eg}^2|\xi_{eg}\rangle|\xi_{eg}\rangle\right)\\
&\qquad+|g\rangle_A|e\rangle_B\left(\beta_{gg}\beta_{ge}|\xi_{gg}\rangle|\xi_{ge}\rangle-i\beta_{eg}\beta_{ee}|\xi_{eg}\rangle|\xi_{ee}\rangle\right)\\
&\qquad+|e\rangle_A|g\rangle_B\left(\beta_{ge}\beta_{gg}|\xi_{ge}\rangle|\xi_{gg}\rangle-i\beta_{ee}\beta_{eg}|\xi_{ee}\rangle|\xi_{eg}\rangle\right)\\
&\qquad+|e\rangle_A|e\rangle_B\left(\beta_{ge}^2|\xi_{ge}\rangle|\xi_{ge}\rangle-i\beta_{ee}^2|\xi_{ee}\rangle|\xi_{ee}\rangle\right)\Big]\\
&=|g\rangle_A|g\rangle_B|\tau_{gg}\rangle+|g\rangle_A|e\rangle_B|\tau_{ge}\rangle+|e\rangle_A|g\rangle_B|\tau_{eg}\rangle+|e\rangle_A|e\rangle_B|\tau_{ee}\rangle,
\end{aligned} \tag{15}$$

where

$$|\tau_{gg}\rangle=\frac{1}{\sqrt{2}}e^{-i\pi/4}\left(\beta_{gg}^2|\xi_{gg}\rangle|\xi_{gg}\rangle-i\beta_{eg}^2|\xi_{eg}\rangle|\xi_{eg}\rangle\right), \tag{16}$$

$$|\tau_{ge}\rangle=\frac{1}{\sqrt{2}}e^{-i\pi/4}\left(\beta_{gg}\beta_{ge}|\xi_{gg}\rangle|\xi_{ge}\rangle-i\beta_{eg}\beta_{ee}|\xi_{eg}\rangle|\xi_{ee}\rangle\right), \tag{17}$$

$$|\tau_{eg}\rangle=\frac{1}{\sqrt{2}}e^{-i\pi/4}\left(\beta_{ge}\beta_{gg}|\xi_{ge}\rangle|\xi_{gg}\rangle-i\beta_{ee}\beta_{eg}|\xi_{ee}\rangle|\xi_{eg}\rangle\right), \tag{18}$$

$$|\tau_{ee}\rangle=\frac{1}{\sqrt{2}}e^{-i\pi/4}\left(\beta_{ge}^2|\xi_{ge}\rangle|\xi_{ge}\rangle-i\beta_{ee}^2|\xi_{ee}\rangle|\xi_{ee}\rangle\right). \tag{19}$$

Alice randomly chooses the SIFT operation or the CTRL operation after she receives the atom from TP. In the meanwhile, Bob also randomly chooses the SIFT operation or the CTRL operation after he gets the atom from TP.

(i) Consider the circumstance that both Alice and Bob choose the SIFT operation. The composite system after Alice and Bob's operations is collapsed into $|x\rangle_A|y\rangle_B|\tau_{xy}\rangle$, where

$x, y \in \{g, e\}$. In order for Eve not being detected in Step 4, $U_F$ should satisfy

$$U_F\left(|x\rangle_A |y\rangle_B |\tau_{xy}\rangle\right) = |x\rangle_A |y\rangle_B |\gamma_{xy}\rangle \ . \tag{20}$$

(ii) Consider the circumstance that Alice chooses the SIFT operation while Bob chooses the CTRL operation. The composite system after Alice and Bob's operations is changed into $|g\rangle_A |g\rangle_B |\tau_{gg}\rangle + |g\rangle_A |e\rangle_B |\tau_{ge}\rangle$ when the measurement result of Alice is $|g\rangle_A$, or $|e\rangle_A |g\rangle_B |\tau_{eg}\rangle + |e\rangle_A |e\rangle_B |\tau_{ee}\rangle$ when the measurement result of Alice is $|e\rangle_A$.

Assume that the measurement result of Alice is $|g\rangle_A$. After Eve performs $U_F$, according to Eq.(20), the global composite system is evolved into

$$U_F\left(|g\rangle_A |g\rangle_B |\tau_{gg}\rangle + |g\rangle_A |e\rangle_B |\tau_{ge}\rangle\right) = |g\rangle_A |g\rangle_B |\gamma_{gg}\rangle + |g\rangle_A |e\rangle_B |\gamma_{ge}\rangle \ . \tag{21}$$

In order for Eve not being detected in Step 4, it needs to satisfy

$$|\gamma_{ge}\rangle = 0 \ . \tag{22}$$

Assume that the measurement result of Alice is $|e\rangle_A$. After Eve performs $U_F$, according to Eq.(20), the global composite system is evolved into

$$U_F\left(|e\rangle_A |g\rangle_B |\tau_{eg}\rangle + |e\rangle_A |e\rangle_B |\tau_{ee}\rangle\right) = |e\rangle_A |g\rangle_B |\gamma_{eg}\rangle + |e\rangle_A |e\rangle_B |\gamma_{ee}\rangle \ . \tag{23}$$

In order for Eve not being detected in Step 4, it needs to satisfy

$$|\gamma_{eg}\rangle = 0 \ . \tag{24}$$

(iii) Consider the circumstance that Alice chooses the CTRL operation while Bob chooses the SIFT operation. The composite system after Alice and Bob's operations is changed into $|g\rangle_A |g\rangle_B |\tau_{gg}\rangle + |e\rangle_A |g\rangle_B |\tau_{eg}\rangle$ when the measurement result of Bob is $|g\rangle_B$, or $|g\rangle_A |e\rangle_B |\tau_{ge}\rangle + |e\rangle_A |e\rangle_B |\tau_{ee}\rangle$ when the measurement result of Bob is $|e\rangle_B$.

Assume that the measurement result of Bob is $|g\rangle_B$. After Eve performs $U_F$, according to Eq.(20) and Eq.(24), the global composite system is evolved into

$$U_F\left(|g\rangle_A |g\rangle_B |\tau_{gg}\rangle + |e\rangle_A |g\rangle_B |\tau_{eg}\rangle\right) = |g\rangle_A |g\rangle_B |\gamma_{gg}\rangle + |e\rangle_A |g\rangle_B |\gamma_{eg}\rangle = |g\rangle_A |g\rangle_B |\gamma_{gg}\rangle . \tag{25}$$

Thus, Eve automatically cannot be detected in Step 4 in this case.

Assume the measurement result of Bob is $|e\rangle_B$. After Eve performs $U_F$, according to Eq.(20) and Eq.(22), the global composite system is evolved into

$$U_F\left(|g\rangle_A |e\rangle_B |\tau_{ge}\rangle + |e\rangle_A |e\rangle_B |\tau_{ee}\rangle\right) = |g\rangle_A |e\rangle_B |\gamma_{ge}\rangle + |e\rangle_A |e\rangle_B |\gamma_{ee}\rangle = |e\rangle_A |e\rangle_B |\gamma_{ee}\rangle \ . \tag{26}$$

Thus, Eve automatically cannot be detected in Step 4 in this case.

(iv) Consider the circumstance that both Alice and Bob choose the CTRL operation. Due to

Eq.(15), the composite system after Alice and Bob's operations is $|g\rangle_A|g\rangle_B|\tau_{gg}\rangle+|g\rangle_A|e\rangle_B|\tau_{ge}\rangle+|e\rangle_A|g\rangle_B|\tau_{eg}\rangle+|e\rangle_A|e\rangle_B|\tau_{ee}\rangle$. After Eve performs $U_F$, according to Eq.(20), the global composite system is evolved into

$$U_F\left(|g\rangle_A|g\rangle_B|\tau_{gg}\rangle+|g\rangle_A|e\rangle_B|\tau_{ge}\rangle+|e\rangle_A|g\rangle_B|\tau_{eg}\rangle+|e\rangle_A|e\rangle_B|\tau_{ee}\rangle\right)$$
$$=|g\rangle_A|g\rangle_B|\gamma_{gg}\rangle+|g\rangle_A|e\rangle_B|\gamma_{ge}\rangle+|e\rangle_A|g\rangle_B|\gamma_{eg}\rangle+|e\rangle_A|e\rangle_B|\gamma_{ee}\rangle. \quad (27)$$

Inserting Eq.(22) and Eq.(24) into Eq.(27), it produces

$$U_F\left(|g\rangle_A|g\rangle_B|\tau_{gg}\rangle+|g\rangle_A|e\rangle_B|\tau_{ge}\rangle+|e\rangle_A|g\rangle_B|\tau_{eg}\rangle+|e\rangle_A|e\rangle_B|\tau_{ee}\rangle\right)$$
$$=|g\rangle_A|g\rangle_B|\gamma_{gg}\rangle+|e\rangle_A|e\rangle_B|\gamma_{ee}\rangle$$
$$=\frac{1}{\sqrt{2}}\left(|\phi^-\rangle_{AB}+i|\phi^+\rangle_{AB}\right)|\gamma_{gg}\rangle+\frac{1}{\sqrt{2}}\left(i|\phi^-\rangle_{AB}+|\phi^+\rangle_{AB}\right)|\gamma_{ee}\rangle$$
$$=\frac{1}{\sqrt{2}}|\phi^-\rangle_{AB}\left(|\gamma_{gg}\rangle+i|\gamma_{ee}\rangle\right)+\frac{1}{\sqrt{2}}|\phi^+\rangle_{AB}\left(i|\gamma_{gg}\rangle+|\gamma_{ee}\rangle\right). \quad (28)$$

In order for Eve not being detected in Step 4, TP's measurement result should always be $|\phi^-\rangle_{AB}$, thus it needs to satisfy

$$|\gamma_{ee}\rangle=-i|\gamma_{gg}\rangle=|\gamma\rangle. \quad (29)$$

(v) Inserting Eq.(29) into Eq.(20), it produces that

$$U_F\left(|g\rangle_A|g\rangle_B|\tau_{gg}\rangle\right)=i|g\rangle_A|g\rangle_B|\gamma\rangle \quad (30)$$

and

$$U_F\left(|e\rangle_A|e\rangle_B|\tau_{ee}\rangle\right)=|e\rangle_A|e\rangle_B|\gamma\rangle\ . \quad (31)$$

Inserting Eq.(22) and Eq.(29) into Eq.(21), it produces

$$U_F\left(|g\rangle_A|g\rangle_B|\tau_{gg}\rangle+|g\rangle_A|e\rangle_B|\tau_{ge}\rangle\right)=i|g\rangle_A|g\rangle_B|\gamma\rangle. \quad (32)$$

Inserting Eq.(24) and Eq.(29) into Eq.(23), it produces

$$U_F\left(|e\rangle_A|g\rangle_B|\tau_{eg}\rangle+|e\rangle_A|e\rangle_B|\tau_{ee}\rangle\right)=|e\rangle_A|e\rangle_B|\gamma\rangle\ . \quad (33)$$

Inserting Eq.(29) into Eq.(25), it produces

$$U_F\left(|g\rangle_A|g\rangle_B|\tau_{gg}\rangle+|e\rangle_A|g\rangle_B|\tau_{eg}\rangle\right)=i|g\rangle_A|g\rangle_B|\gamma\rangle. \quad (34)$$

Inserting Eq.(29) into Eq.(26), it produces

$$U_F\left(|g\rangle_A|e\rangle_B|\tau_{ge}\rangle+|e\rangle_A|e\rangle_B|\tau_{ee}\rangle\right)=|e\rangle_A|e\rangle_B|\gamma\rangle\ . \quad (35)$$

Inserting Eq.(29) into Eq.(28), it produces

$$U_F\left(|g\rangle_A|g\rangle_B|\tau_{gg}\rangle+|g\rangle_A|e\rangle_B|\tau_{ge}\rangle+|e\rangle_A|g\rangle_B|\tau_{eg}\rangle+|e\rangle_A|e\rangle_B|\tau_{ee}\rangle\right)=i\sqrt{2}|\phi^-\rangle_{AB}|\gamma\rangle. \quad (36)$$

Based on Eqs.(30-36), it can be inferred that if Eve want to produce no error in Step 4, the final state of Eve's probe should be independent of not only the operations of Alice and Bob but also their measurement results. As a result, Eve has no way to get $K_A$ or $K_B$.

### 3.2.2 Internal attack

Different from the outside eavesdroppers, internal participants take part in the communication process and have more opportunities to attack, thus compared with the outside attacks, the attacks from internal participants are generally more powerful. [44] Consequently, we should pay more attention on internal attacks. In the following, the internal attack from the semi-honest TP and that from dishonest Alice or Bob are taken into account.

**(1) Attack from the semi-honest TP**

TP cannot collude with other participants due to her semi-honest characteristic. In this protocol, there is a secure key $K_{AB}$ pre-shared between Alice and Bob via SQKD protocol. Although TP can obtain $K_A$ and $K_B$, she still cannot extract $M_A$ and $M_B$, as they are further encrypted with $K_{AB}$.

**(2) Attack from dishonest Alice or Bob**

Alice and Bob are mutually independent and play the identical role in this protocol. Without loss of generality, we only discuss the case where malicious Bob wants to get Alice's private information.

In order to get $K_A$, Bob may launch some attacks, such as the intercept-resend attack, the measure-resend attack, the entangle-measure attack, *et al.*, on the quantum channel between TP and Alice. However, he is inevitably detected as an external attacker during the eavesdropping check process, just as analyzed in section 3.2.1. As a result, Bob has no chance to get $K_A$ without being discovered.

It is possible for Bob to hear $R_A$ from Alice. As $R_A$ is the encrypted result of $K_{AB}$, $M_A$ and $K_A$, in order to decrypt out $M_A$, Bob needs to know $K_A$. However, Bob has no access to $K_A$. Consequently, Bob cannot deduce $M_A$ from $R_A$.

## 4 Simulation based on IBM's Qiskit

In this section, we use IBM's Qiskit to do the simulation experiment to further verify the correctness of the proposed protocol. It is worth noting that any potential eavesdropping attack from an adversary is neglected; each of the following simulation experiment is performed 1024 times; $|g\rangle$ and $|e\rangle$ correspond to 0 and 1, respectively.

As the protocol described, TP prepares the initial product states of $S$ randomly within the four states $\{|gg\rangle, |ge\rangle, |eg\rangle, |ee\rangle\}$, and transmits them into the single-mode cavity. To simulate the evolved states, we use the Hadamard gate (i.e., H gate), the NOT gate, the controlled-NOT gate (i.e., CNOT gate) and the S gate (i.e., $\mathrm{S} = |g\rangle\langle g| + i|e\rangle\langle e|$) to construct the simulation circuit of Fig.3(a). The corresponding simulation results of the evolved states of the four initial states in the form of Bloch spheres are presented in Fig.4.

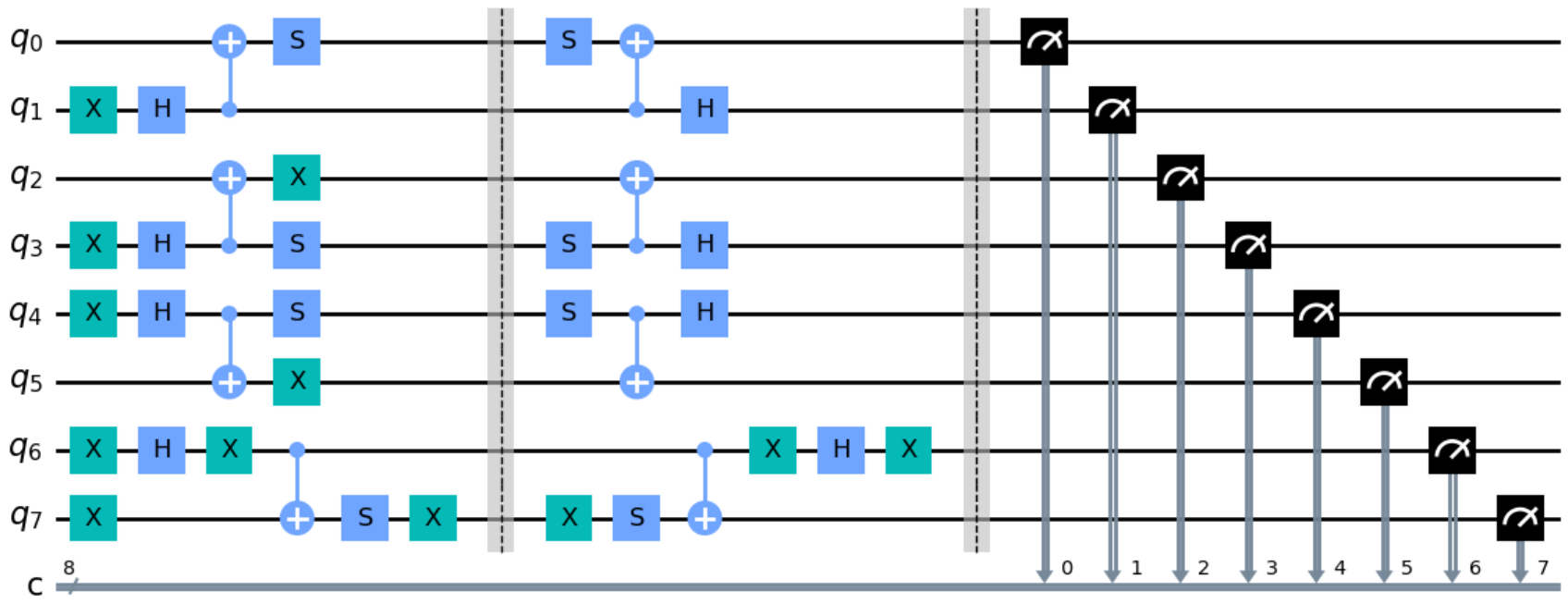

(a) Quantum circuit for the case where both Alice and Bob choose the CTRL operation

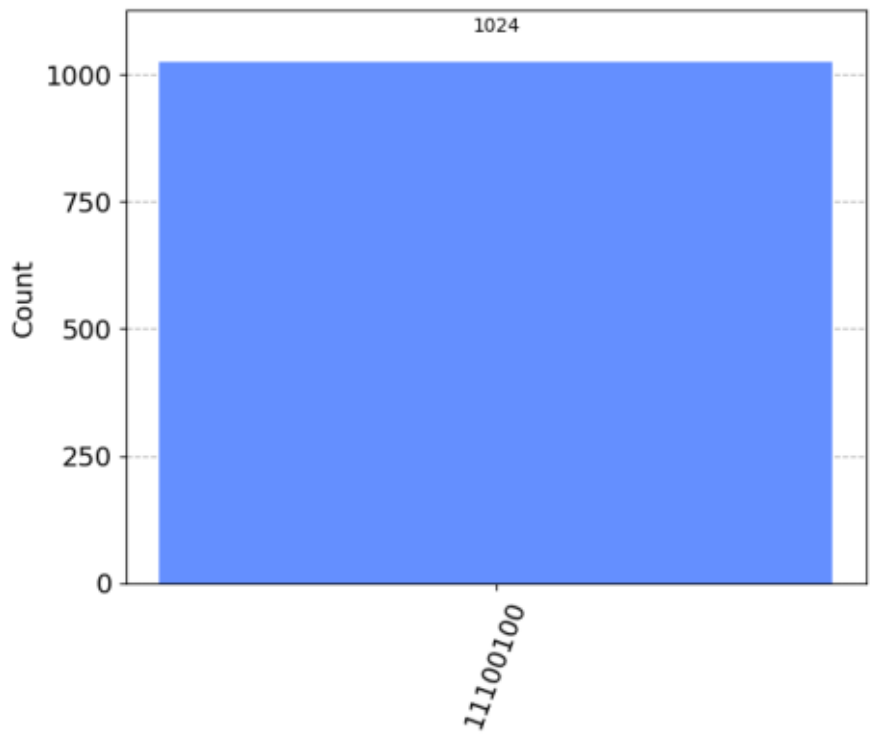

(b) Simulation results of (a)

Fig.3 Quantum circuit and simulation results for the case where both Alice and Bob choose the CTRL operation

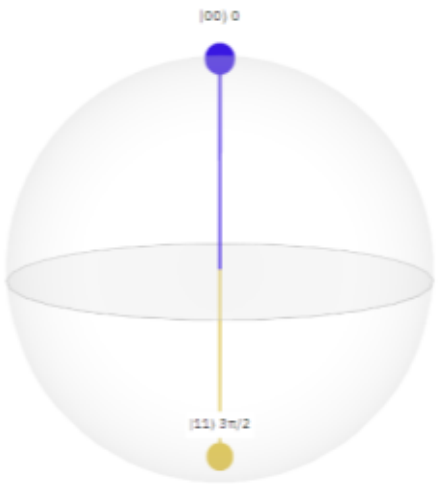

(a) The state of $|gg\rangle$ after evolution

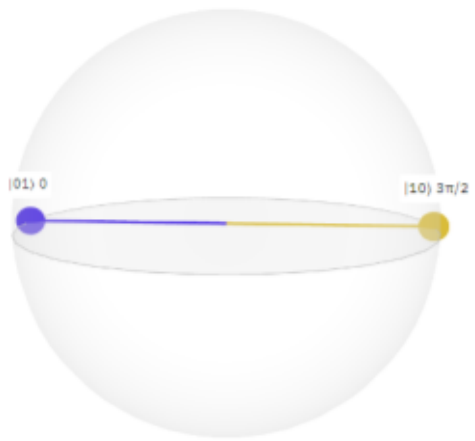

(b) The state of $|ge\rangle$ after evolution

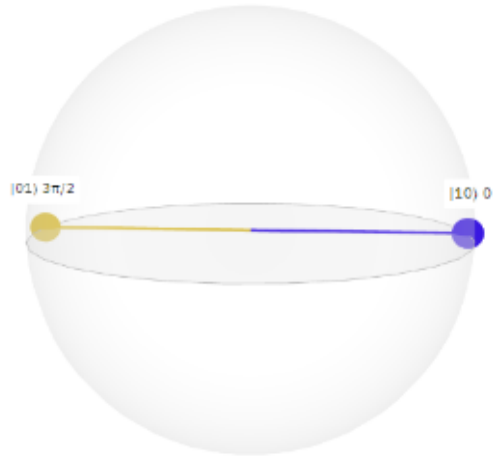

(c) The state of $|eg\rangle$ after evolution

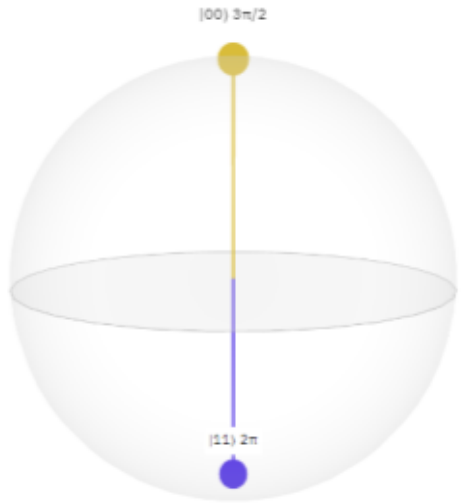

(d) The state of $|ee\rangle$ after evolution

Fig.4 Simulation results of the evolved states of the four initial states in the form of Bloch spheres

Firstly, we focus on the case where both Alice and Bob choose the CTRL operation. The corresponding quantum circuit and simulation results are shown in Fig.3. As shown in Fig.3(a) and 3(b), when the initial states are $|gg\rangle$, $|ge\rangle$, $|eg\rangle$ and $|ee\rangle$, the measurement results obtained by TP correspond to 00, 01, 10 and 11, respectively. In other words, the measurement results of TP are consistent with the evolved states of the initial quantum states.

Furthermore, the corresponding quantum circuit and simulation results for the case where Alice chooses the SIFT operation and Bob chooses the CTRL operation are shown in Fig.5. Here, $q_2$ represents the qubit generated by Alice based on her measurement result. From Fig.5(b), it is obvious that the measurement results of TP are the same as Alice's measure-resend states.

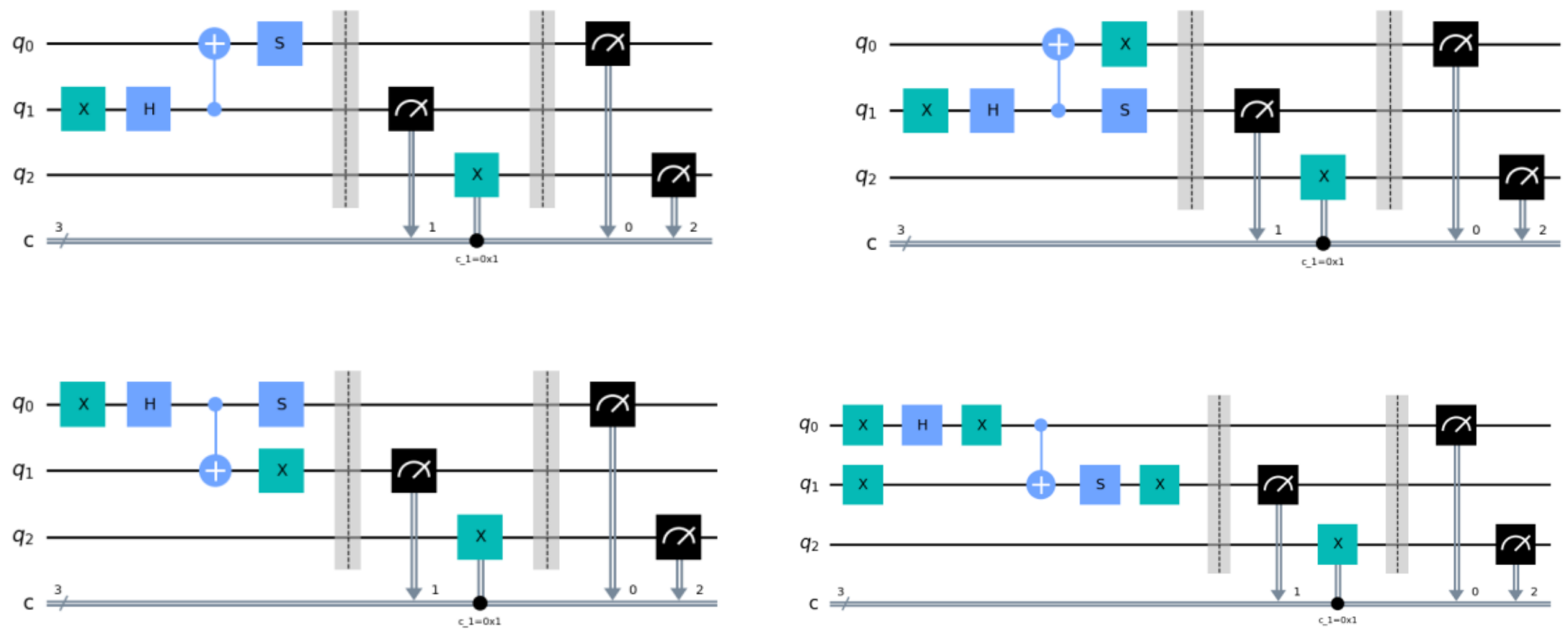

(a) Quantum circuit for the case where Alice chooses the SIFT operation and Bob chooses the CTRL operation

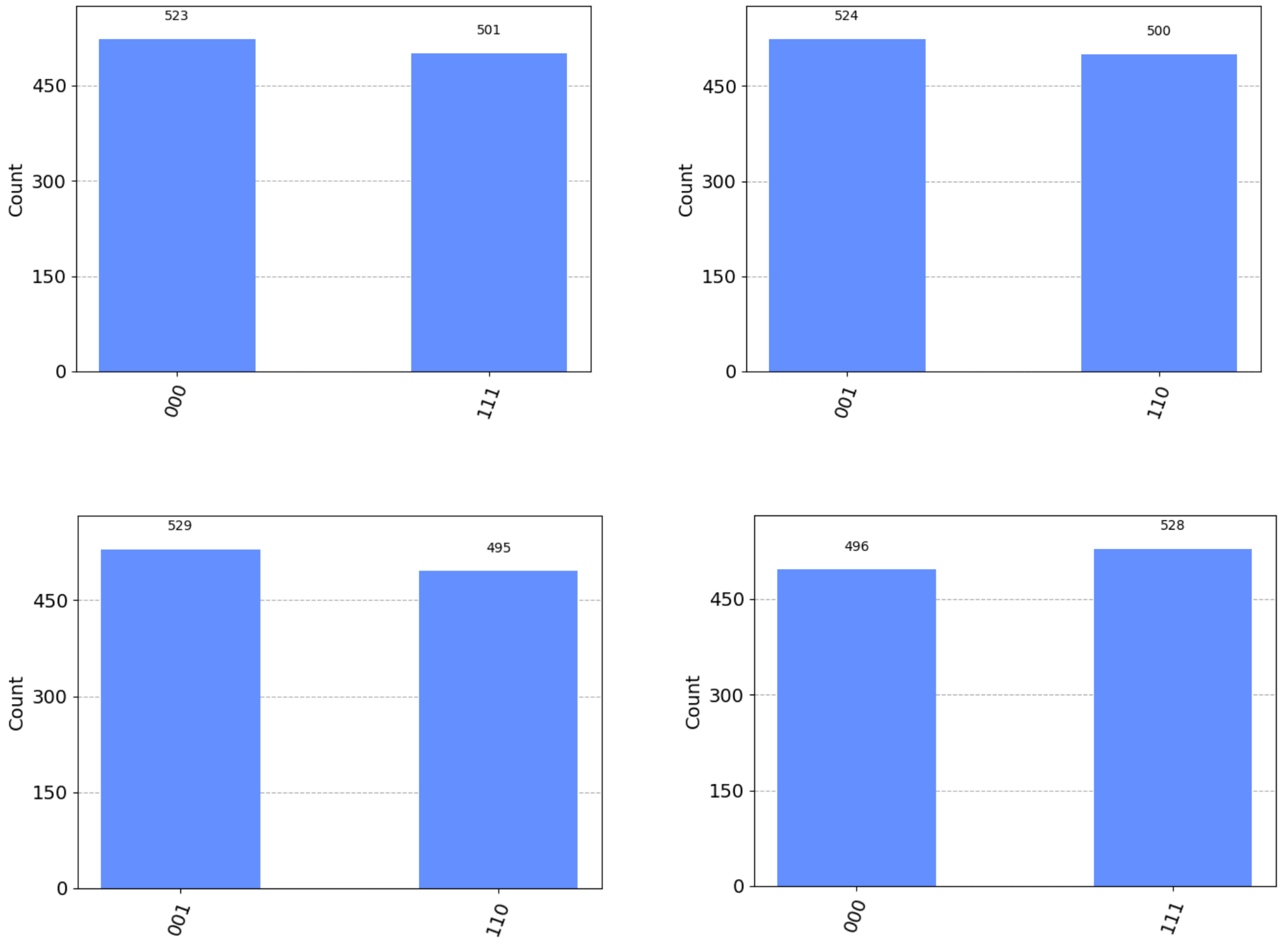

(b) Simulation results of (a)

Fig.5 Quantum circuit and simulation results for the case where Alice chooses the SIFT operation and Bob

chooses the CTRL operation

Next, the corresponding quantum circuit and simulation results for the case where Alice chooses the CTRL operation and Bob chooses the SIFT operation are shown in Fig.6. Here, $q_2$ represents the qubit generated by Bob based on his measurement result. From Fig.6(b), it is obvious that the measurement results of TP are the same as Bob's measure-resend states.

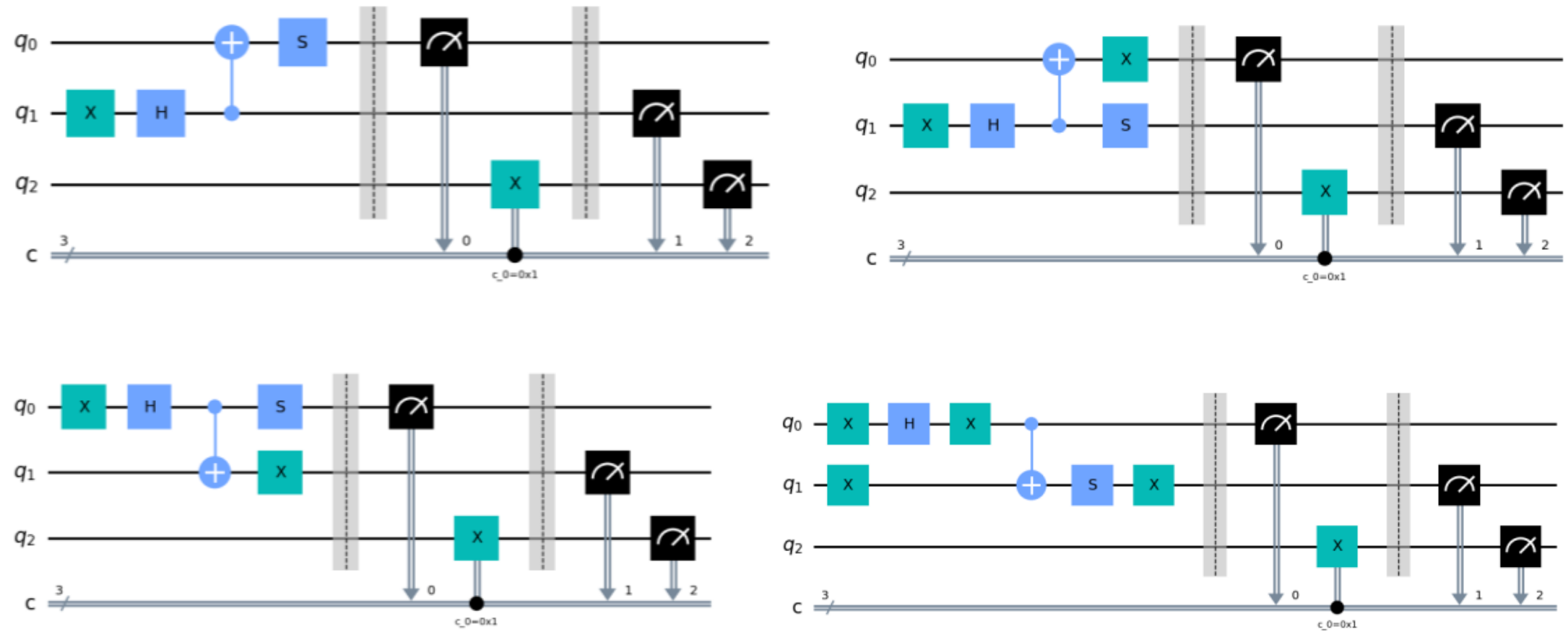

(a) Quantum circuit for the case where Alice chooses the CTRL operation and Bob chooses the SIFT operation

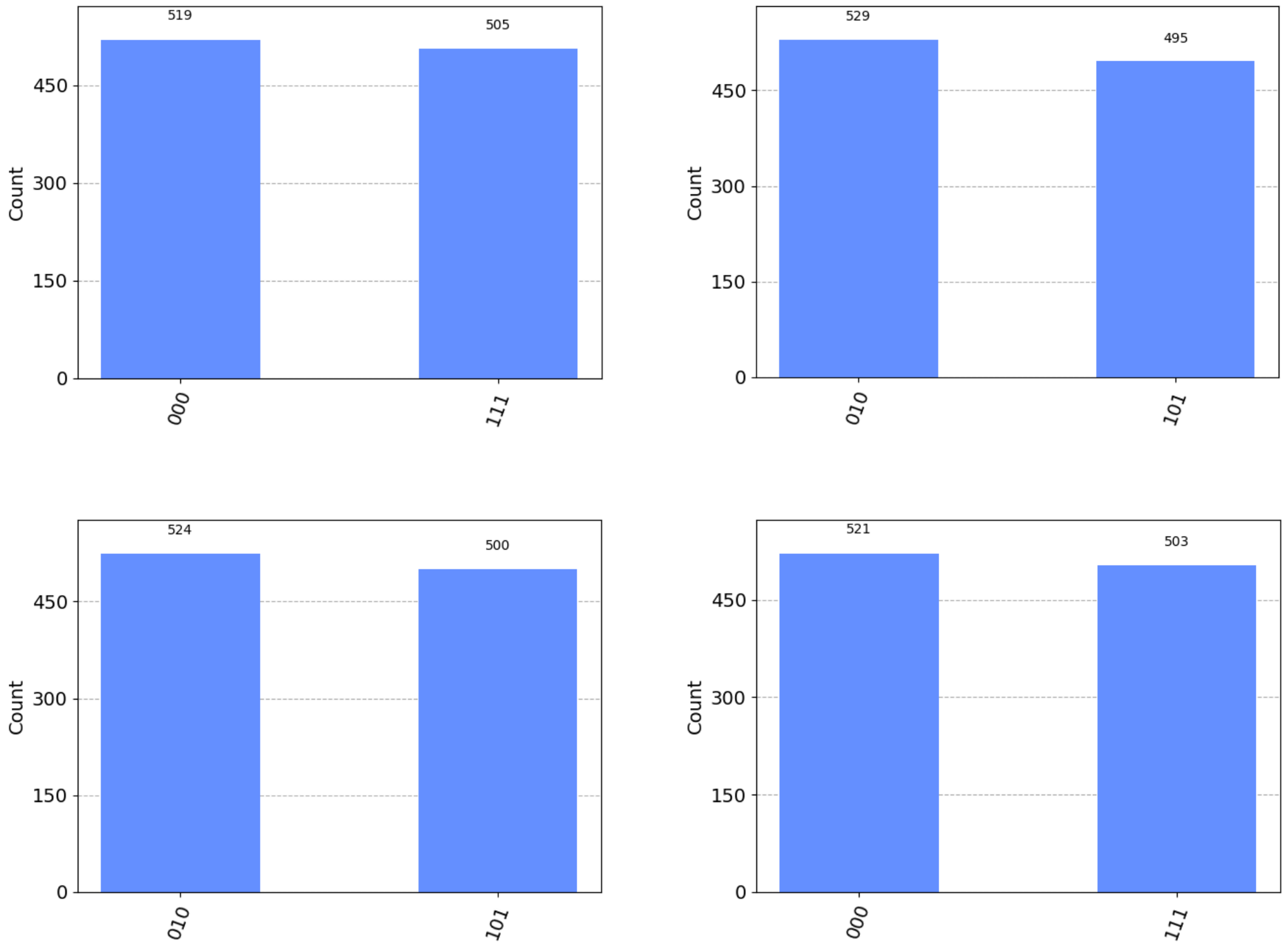

(b) Simulation results of (a)

Fig.6 Quantum circuit and simulation results for the case where Alice chooses the CTRL operation and Bob chooses the SIFT operation

Finally, we discuss the case where both Alice and Bob choose the SIFT operation. The corresponding quantum circuit and simulation results are shown in Fig.7. Here, $q_2$ and $q_3$ represent the qubits generated by Bob and Alice based on their measurement results, respectively. From Fig.7(b), it is no difficulty to see that the measurement results of TP are the same as Alice and Bob's measure-resend states.

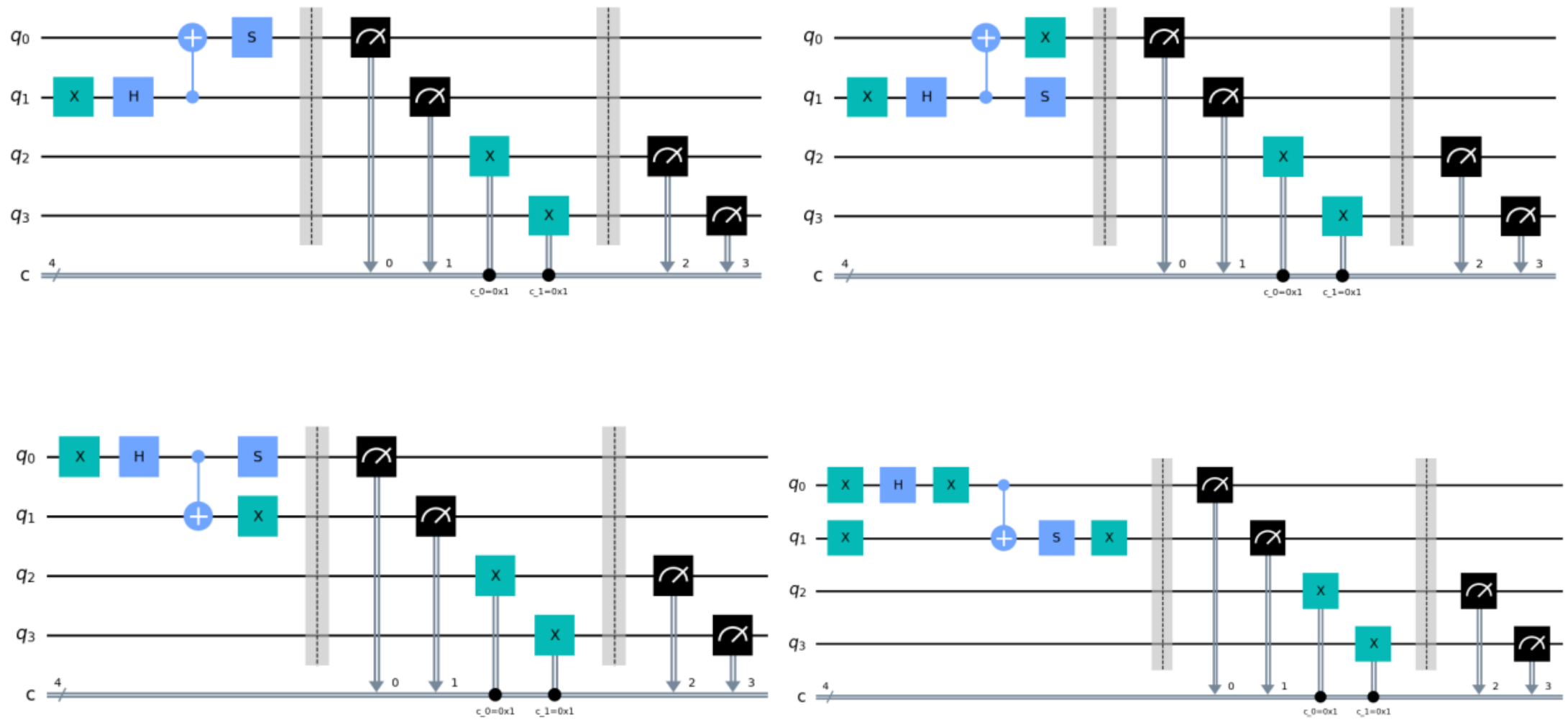

(a) Quantum circuit for the case where both Alice and Bob choose the SIFT operation

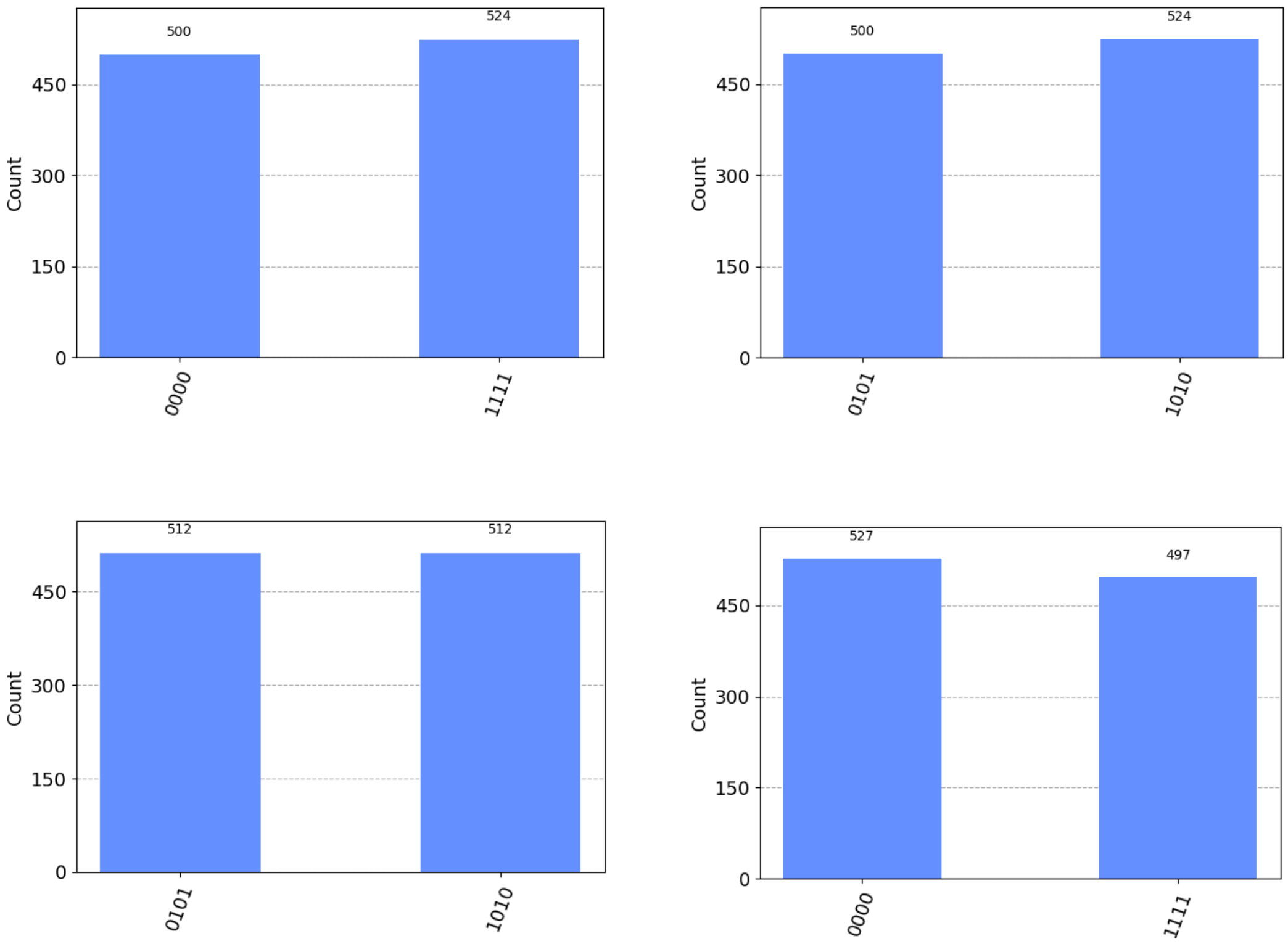

(b) The simulation results of (a)

Fig.7 Quantum circuit and simulation results for the case where both Alice and Bob choose the SIFT operation

In a word, the above simulation results based on IBM's Qiskit show that the proposed SQPC protocol is correct.

## 5 Comparison and conclusion

The communication performance of the proposed SQPC protocol can be evaluated by the qubit efficiency, which is defined as [2]

$$\eta = \frac{\eta_c}{\eta_q + \eta_b}. \tag{37}$$

Here, $\eta_c$, $\eta_q$ and $\eta_b$ denote the number of compared classical bits, the number of consumed qubits and the number of expended classical bits, respectively. The classical bits expended for the eavesdropping check process are disregarded.

In the proposed SQPC protocol, TP prepares $8L$ product states and sends them into the single-mode cavity; afterward, TP sends $S_A$ to Alice and $S_B$ to Bob; when Alice performs the SIFT operations on half of the atoms in $S_A$, she needs to produce $4L$ new atoms; when Bob performs the SIFT operations on half of the atoms in $S_B$, he also needs to produce $4L$ new atoms; and Alice needs to send $R_A$ to TP, while Bob needs to send $R_B$ to TP. In addition, Alice and Bob pre-share $K_{AB}$ via the SQKD protocol of Ref.[41], which expends $24L$ qubits. Thus, we have $\eta_c = L$, $\eta_q = 8L \times 2 + 4L \times 2 + 24L = 48L$ and $\eta_b = L \times 2 = 2L$. As a result, we obtain $\eta = \frac{\eta_c}{\eta_q + \eta_b} = \frac{L}{48L + 2L} = \frac{1}{50}$.

The comparison of the proposed SQPC protocol and other previous SQPC protocols is made in Table 2. According to Table 2, it can be concluded that: (1) as for quantum resource, the proposed SQPC protocol exceeds those of Refs.[18,19,22,26-28], as it doesn't adopt quantum entangled states; (2) as for the usage of quantum entanglement swapping, the proposed SQPC protocol exceeds that of Ref.[18], as it doesn't employ quantum entanglement swapping; (3) as for the usage of delay lines, the proposed SQPC protocol exceeds those of Refs.[19,22], as it doesn't use delay lines; (4) as for qubit efficiency, the proposed SQPC protocol exceeds those of Refs.[18,19, 26,28]; and (5) only the proposed SQPC protocol is realized via cavity QED. In fact, to our best knowledge, up to now, this protocol is the only one which is realized via cavity QED.

Table 2 Comparison of the proposed SQPC protocol and other previous SQPC protocols

| | Feature | Whether is realized via cavity QED | Quantum resources | Type of TP | Usage of quantum entanglement swapping | Usage of unitary operations | Usage of pre-shared key | Usage of delay lines | TP's knowledge about the comparison result | TP's quantum measurements | Qubit efficiency |
|---|---|---|---|---|---|---|---|---|---|---|---|
| The protocol of Ref.[18] | Measure-resend | No | Bell states | Semi-honest | Yes | No | No | No | Yes | Bell basis measurements and single-qubit measurements | $\frac{1}{82}$ |
| The protocol of | Measure-resend | No | Bell states | Semi-honest | No | No | Yes | Yes | Yes | Bell basis measurements and | $\frac{1}{60}$ |
| Ref.[19] | | | | | | | | | | single-qubit measurements | |
| The protocol of Ref.[20] | Measure -resend | No | Two-qubit product states | Semi-honest | No | No | Yes | No | Yes | Single-qubit measurements | $\frac{1}{34}$ |
| The protocol of Ref.[21] | Measure -resend | No | Single qubits | Semi-honest | No | No | No | No | Yes | Bell basis measurements and single-qubit measurements | $\frac{1}{50}$ |
| The second protocol of Ref.[22] | Measure-Randomization -resend | No | Bell states | Semi-honest | No | No | Yes | Yes | Yes | Bell basis measurements | $\frac{1}{32}$ |
| The protocol of Ref.[26] | Measure -resend | No | Bell states | Semi-honest | No | No | Yes | No | Yes | Bell basis measurements | $\frac{1}{58}$ |
| The protocol of Ref.[27] | Measure -resend | No | W states | Semi-honest | No | No | Yes | No | Yes | Bell basis measurements and single-qubit measurements | $\frac{1}{42}$ |
| The protocol of Ref.[28] | Measure -resend | No | Bell states | Semi-honest | No | No | Yes | No | Yes | Bell basis measurements and single-qubit measurements | $\frac{1}{70}$ |
| The proposed protocol | Measure -resend | Yes | Two-atom product states | Semi-honest | No | No | Yes | No | Yes | Two-atom entangled state measurements and single-atom measurements | $\frac{1}{50}$ |

In conclusion, in this paper, we propose a novel measure-resend SQPC protocol which is realized via cavity QED by making use of the evolution law of atom. With the help of a semi-honest TP, the proposed protocol can compare the equality of private inputs from two semiquantum parties who only own limited quantum capabilities. The proposed protocol uses product states as initial quantum resource which are much easier to prepare than entangled states. In addition, the proposed protocol employs none of unitary operations, quantum entanglement swapping operation or delay lines. Security proof validates that the proposed protocol can resist both the external attack and the internal attack. This protocol is actually the only SQPC protocol at present which is realized via cavity QED.

According to Refs.[15,16], a semiquantum cryptography scheme has three different security levels: completely robust, partly robust and completely nonrobust. Complete robustness implies that if Eve wants to get nonzero information about the INFO string, she will introduce errors on the checked bits with a nonzero probability. The complete robustness of the proposed SQPC scheme has been verified in detail in (3) of Sect.3.2.1. Due to limit of space, we leave the proof of security from the viewpoint of quantum information theory in the future.

**Acknowledgments**

Funding by the National Natural Science Foundation of China (Grant No.62071430) and the Fundamental Research Funds for the Provincial Universities of Zhejiang (Grant No.JRK21002) is gratefully acknowledged.